\documentclass[amsmath,amssymb,onecolumn,superscriptaddress]{article}
\usepackage{graphicx}
\usepackage{amsfonts}
\usepackage{color}
\usepackage{amsmath}
\usepackage{color}

\usepackage{authblk}
\usepackage{framed}
\definecolor{shadecolor}{rgb}{1,0.8,0.3}
\usepackage[utf8x]{inputenc}
\usepackage{amsmath}
\usepackage{placeins}
\usepackage[textwidth=17.31cm]{geometry}
\usepackage[super]{natbib}
\usepackage{abstract}

\def\<{\langle}
\def\>{\rangle}
\def\tw{t_w}
\def\twf{t_p}

\title{Dynamic phase coexistence in glass--forming liquids}
\author[1,*]{Raffaele Pastore} 
\author[1]{Antonio Coniglio}
\author[2,1]{Massimo Pica Ciamarra}
\affil[1]{
CNR--SPIN, Dipartimento di Scienze Fisiche,
Universit\'a di Napoli Federico II, Italy
}
\affil[2]{
Division of Physics and Applied Physics, School of Physical and Mathematical Sciences, \newline Nanyang Technological University, Singapore
}
\affil[*]{Correspondence and requests for materials should be addressed to R.P. (pastore@na.infn.it)}

\date{}
\begin{document}
    \maketitle
\begin{abstract}

One of the most controversial hypotheses for explaining the heterogeneous dynamics of glasses 
postulates the temporary coexistence of two phases characterized by a high and by a low diffusivity. 
In this scenario, two phases with different diffusivities coexist for a time of the order of the relaxation time and mix afterwards. 
Unfortunately, it is difficult to measure the single-particle diffusivities to test this hypothesis. Indeed, 
although the non-Gaussian shape of the van-Hove distribution suggests the transient existence of a diffusivity distribution, 
it is not possible to infer from this quantity whereas two or more dynamical phases coexist. 
Here we provide the first direct observation of the dynamical coexistence  of two phases with different diffusivities, 
by showing that in the deeply supercooled regime the distribution of the single-particle diffusivities acquires a transient 
bimodal shape. We relate this distribution to the heterogeneity of the dynamics and to the breakdown of the Stokes-Einstein relation, 
and we show that the coexistence of two dynamical phases occurs up to a timescale growing faster than the relaxation time on cooling, 
for some of the considered models. 
Our work offers a basis for rationalizing the dynamics of supercooled liquids and for relating their structural and dynamical properties.
\end{abstract}

\section*{Introduction}
Glass forming systems have a spatially and temporally heterogeneous dynamics\cite{DHbook}
as revealed, for instance, by the time evolution of the Van Hove (vH) distribution function.
This is the probability distribution that a particle has moved of a distance $r$
along a fixed direction at time $t$, and is a Gaussian
with variance $Dt$ if particles move with a constant diffusion
coefficient $D$. Conversely, in glass formers the vH distribution has a temporary
non-Gaussian shape\cite{Kob97,Donati98,Donati99, Kob07}, that indicates the temporary coexistence of particles with different
diffusion coefficients.
It has been suggested\cite{Schmidt1991,Glotzer,Ediger,Chandler_Science, Cammarota2010, Tanaka}
that this dynamical heterogeneity reflects the transient coexistence of two phases with different dynamical
features, commonly indicated as the `fast' and as the `slow' phase. 
However, in equilibrium systems it has not yet been identified a dynamical order parameter with a transient
bimodal probability distribution, which would support the existence of two coexisting phases; 
indeed, up to now a dynamical order parameter with a bimodal distribution has only been identified in structural glasses driven 
out of equilibrium introducing a field pinning some of the particles\cite{Chandler_Science},
and thus inducing the two phases, or more complex constraints on the relaxation dynamics\cite{Cammarota2010}. 
Because of this, in equilibrium supercooled liquids the `fast' and the `slow' phase are usually empirically defined, for instance by 
considering as `fast' $5\%$ particles, chosen to be the ones with the largest displacement\cite{WeeksScience}.
These empirical criteria are used because the vH distribution cannot have a bimodal or multimodal shape
allowing for the clear identification of different coexisting phases.
Indeed, if phases with different diffusivities coexist,
then the vH distribution will be the weighted sum of different Gaussian functions, all centered in $r = 0$,
and will thus have a single maximum. 
This clarifies that, in order to investigate whereas two or more dynamical phases coexist, one should
investigate the diffusivity distribution, not the vH distribution. 
Unfortunately, the investigation of the diffusivity distribution is difficult,
and only stationary diffusivity distributions can be obtained via a direct inversion of the vH distribution 
\cite{Granick, Sengupta}.

Here we report the first measure of the time evolution of the single particle diffusion coefficient,
for different model systems: the standard Kob--Andersen Lennard--Jones (KALJ) binary mixture\cite{KA1,KA2,KA3},
a binary mixture of soft-spheres in two dimensions\cite{PastoreSM}, and the Kob--Andersen lattice 
gas model\cite{KAKC,PRL2011}.
In the deeply supercooled regime, we find this distribution to temporarily acquire a bimodal shape, 
thus proving the transient coexistence of two distinct dynamical phases.
In the long--time limit the two phases mix and the diffusivity distribution acquires
the expected Gaussian shape, with a variance to mean ratio we show
to be related to the breakdown of the Stokes--Einstein relation. 

We succeeded in measuring the single particle diffusion coefficient by exploiting the intermittent nature of the single particle motion in structural 
glasses\cite{Inter1,Inter2,Inter3,Inter4, Inter5, Makse2009, SM15}. Indeed,
particles in a glass spend most of their time confined within the
cages formed by their neighbors, seldom hopping to different cages.
This allows to describe the dynamics through the continuous time random walk (CTRW) formalism, reviewed in the Appendix.
In this framework, the diffusivity of each particle at a given time is proportional to its number
of jumps, and the distribution of diffusivities
is equivalent to the distribution of the number of jumps per particle.
Accordingly, in the following we first show that the CTRW approach 
quantitatively describes the dynamics of the considered systems, 
when cages and jumps are identified 
using a recently developed parameter--free algorithm\cite{PastoreSM}.
We discuss in detail the KALJ system to show that the CTRW approach
quantitatively describes the relaxation
dynamics of atomistic systems, not only of kinetic lattice models
\cite{berthier_epl2005, Chandler_SE}.
Then, we use this approach to measure the diffusivity distribution
and to investigate its time evolution.
\section*{Results}
\subsection*{CTRW description of the dynamics}
\begin{figure}[t!!]
\begin{center}
\includegraphics*[scale=0.33]{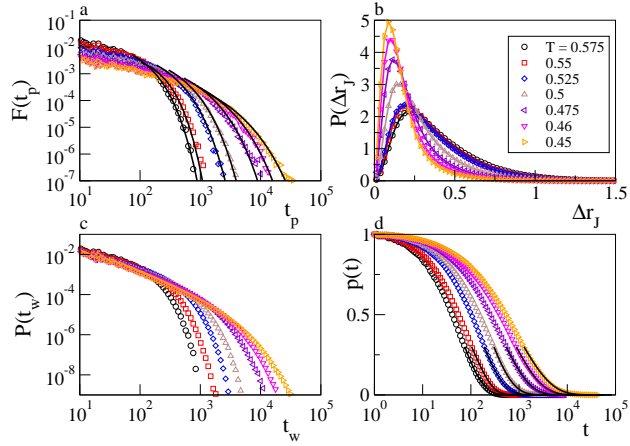}
\end{center}
\caption{\label{fig:p_njumps}
Persistence and cage--jump properties.
Panels a,b and c show the probability distributions of the persistence
time $\twf$, of the jump length $\Delta r_J$, and of the waiting time $\tw$.
Panel d illustrates the decay of the persistence. Full lines in panel d are fits to stretched exponentials, 
while those in panel a are the corresponding predictions for $F(\twf)$ (see text).
All data refer to species $a$ of the KALJ mixture.
Analogous results for species $b$ are shown in Fig.S1.
}
\end{figure}
In order to prove that the CTRW approach provides a quantitative description
of the dynamics of the KA mixture, we have performed a careful analysis
of the single particle cage--jump intermittent motion,
for temperatures slightly above the mode--coupling one\cite{KA1,KA2,KA3}, 
$T_{\rm mct} \simeq 0.435$.
Fig. 1a--c illustrate the distribution of
the persistence time $F(\twf)$ and of the jump length $P(\Delta r)$,
that fix the temporal and spatial features of the system in the CTRW approach, 
as well as the distribution of the time particles wait in their cages
before making a jump, $P(\tw)$.
No correlations between the persistence time and the jump length have been found,
in agreement with the CTRW scenario.
Panel d illustrates the decay of the persistence.
At short times all jumps contribute to the decay of the persistence;
we therefore observe $p(t)=1-t/\<\tw\>$, as $\<\tw\>^{-1}$
is the rate at which particles jump, and $F(\twf) = -dp(t)/dt = \<\tw\>^{-1}$. 
At long times ($t \simeq t_p$) the persistence is found to decay with a 
stretched exponential, $p(t) \propto \exp\left(-(t/\tau)^\beta\right)$.
This implies $F(\twf) = -dp(t)/dt \propto \tau^{-\beta} t^{\beta-1} \exp \left(-(t/\tau)^\beta 
\right)$ as verified in Fig.\ref{fig:p_njumps}a.
\begin{figure}[t!]
\begin{center}
\includegraphics*[scale=0.33]{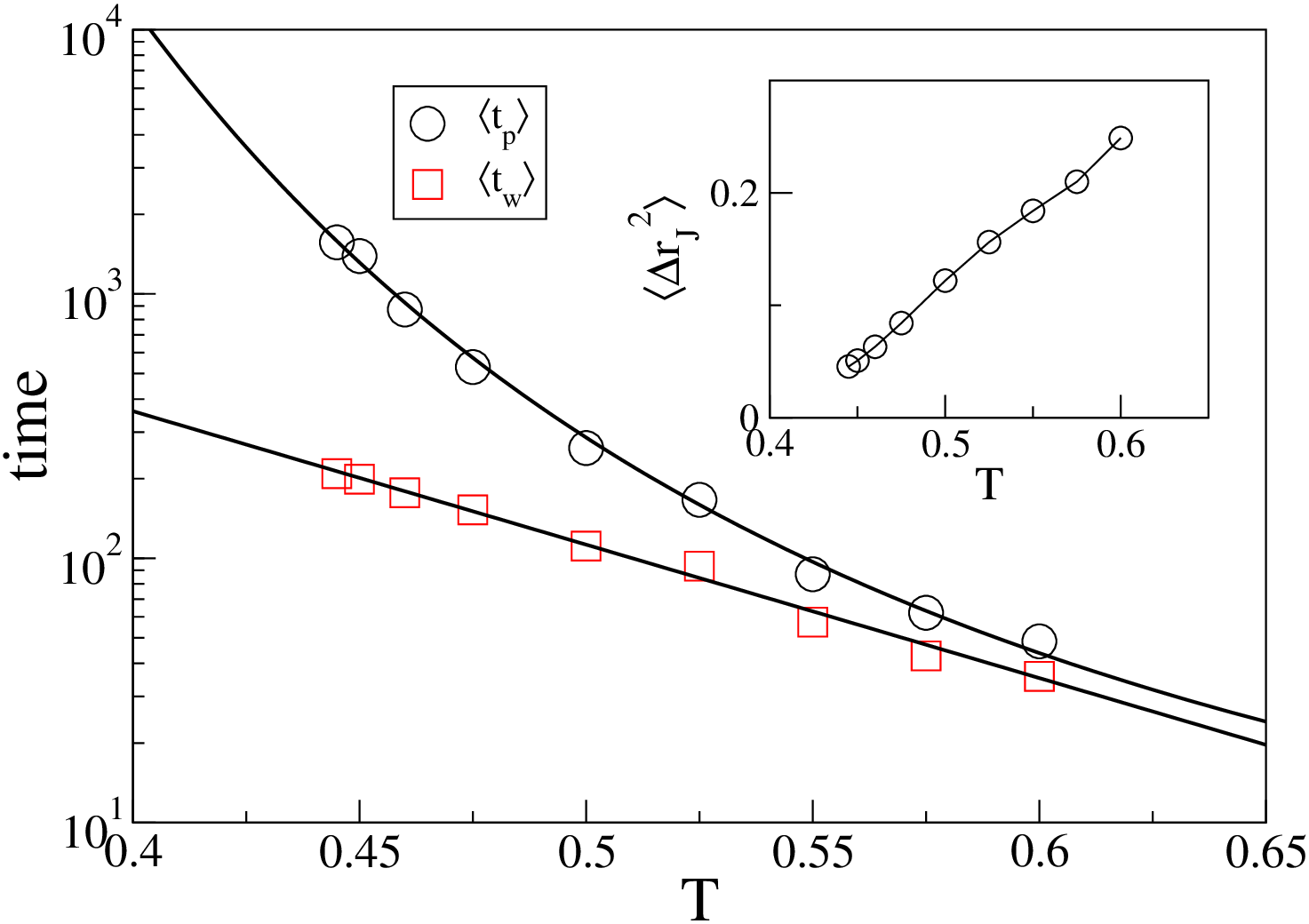}
\end{center}
\caption{\label{fig:times}
Cage--jump time and length scales.
Temperature dependence of the average time particles persist in a cage before 
making the first jump, $\<\twf\>$,
and of the average cage residence time, $\<\tw\>$. 
$\<\tw\>$ is well described by an Arrhenius $\<\tw\> \propto \exp\left(A/T\right)$ (full line).
$\<\twf\>$ grows \'a super--Arrhenius law. The dashed line is a fit to
$\<\twf\> \propto \exp\left(A/T^B\right)$, with $B = 2.4$, but other functional forms, including the Vogel--Fulcher one,
also describe the data.
The inset illustrates the temperature dependence of the average jump length. The line is a guide to the eye.
All data refer to species a of the KA LJ mixture.
Analogous results for species b are shown in Fig.S2.
}
\end{figure}

\begin{figure}[t!]
\begin{center}
\includegraphics*[scale=0.39]{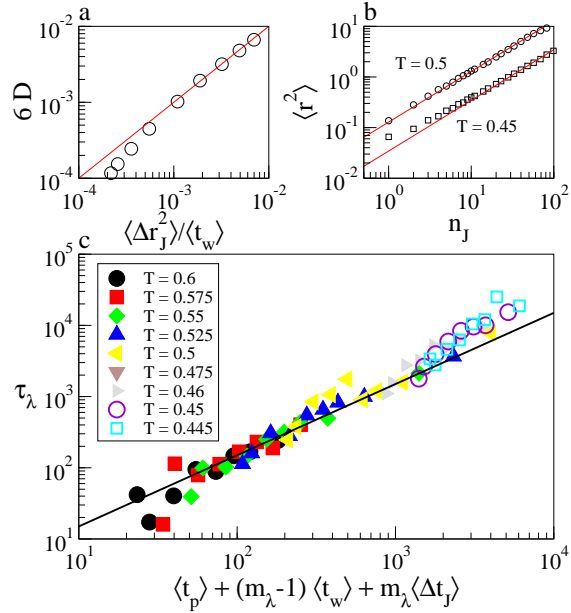}
\end{center}
\caption{\label{fig:tau_a}
Structural relaxation and cage--jump properties.
The diffusivity (panel a) and the relaxation time 
at a generic length scale $\lambda$ (panel c) versus
their predictions in the CTRW approach. Small deviations
are observed at the lowest temperatures due 
to the emergence of a subdiffusive transient
in the dependence of the mean square displacement on the number
of jumps, as in panel b at $T = 0.45$. This indicates 
that successive jumps of a same particle becomes spatially correlated.
All data refer to species $a$ of the KA LJ mixture.
Analogous results for species $b$ are shown in Fig.S3.
}
\end{figure}

The temperature dependence of the main quantities characterizing the cage--jump motion
is illustrated in Fig.\ref{fig:times}. We observe 
the time scales $\<\tw\>$ and $\<\twf\>$ to have an Arrhenius
and a super--Arrhenius behavior, respectively, and the average squared jump
length to decrease on cooling. The temperature dependence of these quantities can be used to 
rationalize those of the diffusion coefficient $D$
and of the structural relaxation time $\tau_\lambda$ at different
wavelength $\lambda$ (wavevector $2\pi/\lambda$), which
are commonly accessed experimentally. Indeed, in the
CTRW approach it is easy to verify that $D = 6 \< \Delta r_J^2 \> /\<t_w\>$. 
Fig.\ref{fig:tau_a}a illustrates that this relation is verified
at the highest temperatures. Deviations emerge on cooling as subsequent jumps
of a same particle becomes spatially correlated, as clarified by the 
subdiffusive dependence of the mean square displacement versus the number of jumps
illustrated in panel {\it b}.
The relaxation time $\tau_\lambda$ scales as the average time a particle needs to move a distance $\lambda$.
Since in the CTRW approach subsequent jumps of a same particle are spatially uncorrelated, 
this time is that particles need to perform, on average, $m_\lambda(T) = 
\lambda^2/\<\Delta r^2_J(T)\>$ jumps, and is fixed by the average time particles wait before
making the first jump, $\< \twf \>$, and the subsequent ones, $\< \tw \>$, as well
as by the average jump duration, $\<\Delta t_J\>$:
\begin{equation}
\label{eq:t_l}
\tau_\lambda \propto \<\twf\> + (m_\lambda-1)\<\tw\> + m_\lambda\<\Delta t_J\>.
\end{equation}
The last term is actually negligible at low temperatures, where $\<t_w\> \gg 
\<\Delta t_J\>$\cite{PastoreSM}.
Fig.\ref{fig:tau_a} shows that this prediction
agrees very well with the measured data in the investigated range of $\lambda$, with a coefficient
of proportionality of the order of $1$. As in the case of the diffusivities, small deviations
are observed at the lowest temperatures.
These results clearly demonstrate that $\<t_w\>$ and $\<\twf\>$
respectively correspond to the $\beta$ and to the $\alpha$ relaxation time scales of structural glasses\cite{berthier_epl2005, Debenedetti,  Hedges2007},
and confirm that the breakdown of the Stokes--Einstein (SE) relation, which is the increase of the product $\tau_\lambda D$ on cooling,
is mainly due to the increase of the $\<\twf\>/\<\tw\>$ ratio, as in lattice model, but it is also
affected by the temperature dependence of the jump length.
Indeed, we find the length scale\cite{berthier_epl2005}
below which the breakdown of the SE relation occurs, 
$\lambda \simeq
\left[ \<\Delta r^2_J(T)\> \left( 1 + \<\tw\>/\<\twf\>\right) \right]^{1/2}$,
that is estimated equating the first two terms of the r.h.s. of Eq.\ref{eq:t_l},
to also depend on the spatial features of the jumps.

\subsection*{Diffusivity distribution}
\begin{figure}[t!]
\begin{center}
\includegraphics*[scale=0.36]{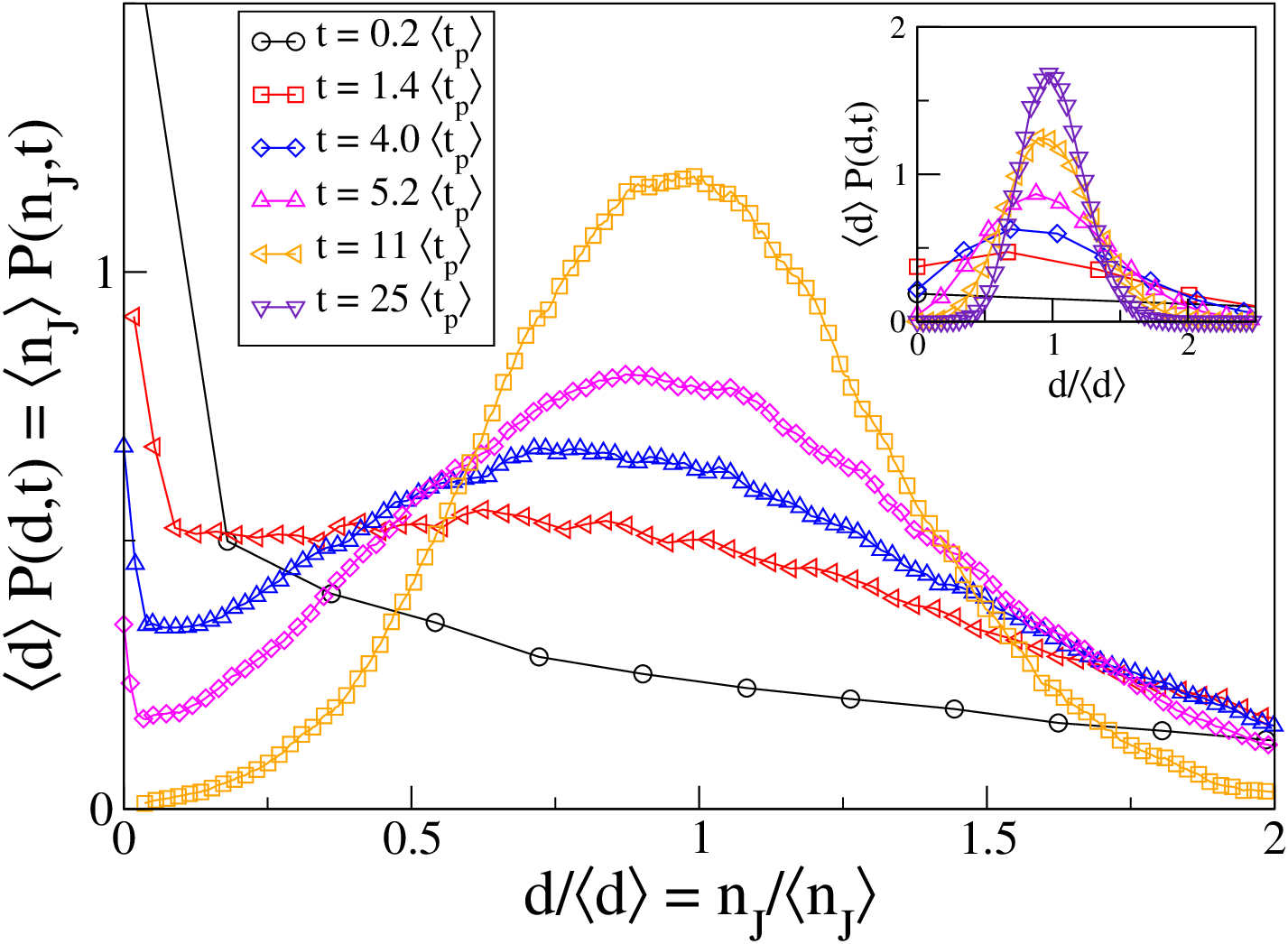}
\end{center}
\caption{\label{fig:panel1}
Diffusivity distribution.
Probability distribution of the single particle diffusion coefficient at different
time, rescaled by the average diffusivity, at $T=0.6$ and  $t= 0.2, 1.4, 4, 5.2, 11, 25 \<t_p\>/\<t_w\>$ with $\<t_p\>/\<t_w\>\simeq 1$ (inset) ,
and at $T=0.45$ and  $t= 0.65, 4.3, 7.7, 15, 29 \<t_p\>/\<t_w\>$ with $\<t_p\>/\<t_w\>\simeq 10$ (main panel).
At low temperature and intermediate time, the distribution acquires a temporary bimodal shape 
with the maxima occurring at $d/\<d\>=0$ and $d/\<d\> \simeq 1$, respectively.
All data refer to species $a$ of the KA LJ mixture.
Analogous results for species $b$ are shown in Fig.S4.
}
\end{figure}

The quantitative description of the relaxation
dynamics through the statistical features of the cage--jump motion
allows to exploit the features of the CTRW approach 
to measure the distribution of the single
particle diffusivities. Indeed, within the CTRW 
the diffusivity
of particles that have performed $n_J$ jumps
at time $t$ is $d(n_J,t) = n_J(t) \<\Delta r^2_J\>/6t$.
The diffusivity is therefore simply proportional to the number of jumps per unit time.
Fig.s\ref{fig:panel1} illustrates the time
evolution of the number of jumps per particle rescaled by the average number of jumps
$\<n_J(t)\> = t/\<\tw\>$, which coincides with the distribution
of the single particle diffusion coefficient
normalized by the average diffusion coefficient, $\<d\> = D$.
The inset and the main panel show
results obtained at a high and at a low temperature, respectively.
For $t/\<\tw\> \ll 1$, $P(d;t)$ is peaked around zero as most
particles have not jumped; conversely, in the infinite
time limit the distributions have a Gaussian shape with average value
$\<d\>$. We observe that, at high temperature,
the distribution gradually broadens in time,
and its maximum move from $d \simeq 0$ to $d \simeq \<d\>$.
At low temperature, conversely,
the distribution acquires a temporary bimodal shape
before reaching the asymptotic Gaussian one. 
The bimodal shape proves the existence of a time window
in which two phases of particles with different mobilities coexist.
The two phases emerge because of the existence of two well separated timescales $\<\twf\>$ and $\<\tw\>$.
Indeed the slow timescale, $\<\twf\>$, controls the value of the peak at $d = 0$, that equals the persistence correlation function, $P(d=0;t) = p(t)$.
Conversely, the fast timescale, $\<\tw\>$, controls the average value of the
distribution, as the position of the second maximum asymptotically occurs at 
$d = \<d\> \propto t/\<\tw\>$.
We stress that the phases with an high and with a low diffusion
coefficient cannot be uniquely associated to particles
that have moved over a small or over a large distance, respectively,
as the average displacement of each particle is zero.
This is why a bimodal distribution is not observed in the vH distribution function.

\begin{figure}[t!]
\begin{center}
\includegraphics*[scale=0.33]{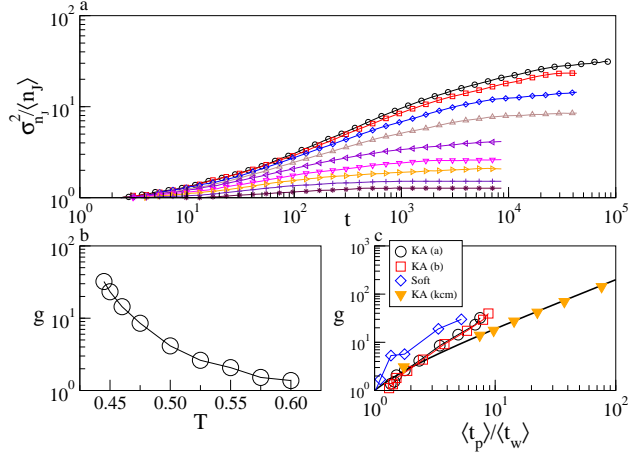}
\end{center}
\caption{\label{fig:panelg}
Variance to mean ratio of the distribution of the number of jumps per particle.
Time evolution (a) of the variance to mean ratio of the distribution of the number of jumps per particle,
and temperature dependence of its asymptotic value (b). Data refer to species $a$.
Analogous results for species $b$ are reported in Fig.S4. 
Panel c illustrates that, in the deeply supercooled
regime, the asymptotic value scales as 
$g \propto \<\twf\>/\<\tw\>$, for both components of
the KA mixture and for other model systems (see text).
The full line is the CTRW prediction, $g = 2\<\twf\>/\<\tw\> - 1$.
}
\end{figure}

The time evolution of the distribution of the diffusivities gives further insights
into the dynamics of the system. Indeed, Fig.s\ref{fig:panelg}a,b
show that at long times the variance to mean ratio of $P(n_J;t)$ reaches a 
plateau value $g = \sigma^2_{n_J}/\<n_J\>$, that grows on cooling.
This plateau value can be related to the ratio of the two timescales $\<\twf\>$ and $\<\tw\>$.
In fact, within the CTRW framework\cite{Feller49,FellerBook}
$\<n_J\> = t/\<\tw\>$ and
$\sigma^2_{n_J} = t \sigma^2_{\tw}/\<\tw\>^3$, where 
$\sigma^2_{\tw} = \<\tw^2\>-\<\tw\>^2$. 
Given the relation between the persistence time and the waiting time distributions\cite{Lax} (see Appendix),
it follows $\sigma^2_{\tw} = 2 \<\tw\> \<\twf\> - \<\tw\>^2$ and thus $g = 2 \<\twf\>/\<\tw\>  - 1$. 
We have verified this prediction considering, beside the KA model, 
also a binary mixture of harmonic spheres\cite{PastoreSM} and the kinetically constrained
Kob--Andersen three dimensional lattice gas model\cite{KAKC,PRL2011},
as illustrated in Fig.\ref{fig:panelg}(inset). 
The lattice model confirms our predictions. The molecular dynamics
simulations reproduce the asymptotically proportionality between $g$ and $\<\twf\>/\<\tw\>$,
even though there are small deviations with respect to the CTRW prediction,
suggesting the emergence of correlations between successive
waiting times at low temperature. 

A physical interpretation of the proportionality
between $g$ and $\<\twf\>/\<\tw\>$ 
is obtained considering that a distribution with the long--time 
features of $P(n_J;t)$, i.e. a 
Gaussian distribution with variance $\sigma^2_{n_J} = \<n_J\>g$,
is obtained by randomly assigning the jumps to the particles, in group 
of $g$ elements. Consistently, at high temperature
$g = 1$ and $P(n_J;t)$ corresponds to that 
obtained by randomly assigning the jump to the particles,
i.e. a Poisson distribution. The increase of $g$ on cooling indicates
that at low temperature one might observe, in the same time interval, some particles
to perform $g$ jumps, and other particles to perform no jumps at all,
which clearly suggests $g \propto \<\twf\>/\<\tw\>$.

\subsection*{Spatial correlations}
The CTRW approach does not make any assumption about the spatial correlations
between the jumps of different particles.
However, particularly in a facilitation
scenario in which the jump of a particle facilitates the jumps of nearby particles,
one expects these correlation to exist, and hence the two dynamical
phases to be spatially segregated.
Previous investigations
of the spatio--temporal heterogeneities of structural glasses~\cite{DHbook}
also suggest that this should be the case.
Here we investigate these spatial correlation focusing on two correlation functions,
both of them related to a scalar field associated to the number of jumps,
$n_J(r,t) = 1/N \sum_{i}^N n_J^{(i)}(t) \delta(r-r_i)$. Note that
$n_J(r,t)t dr$ is proportional to the average diffusion coefficient
of the particles in the volume element $dr$.

First, we consider the spatial correlations between the particles that
have not jumped at time $t$, 
\begin{equation}
c_0(r,t) = \frac{\langle \delta(n_J(0,t))\delta(n_J(r,t))\rangle - \langle p\rangle^2}{\langle p^2\rangle - \langle p\rangle^2}.
\end{equation} 
This equals the correlation function of the particles that have jumped, and thus of the particles
that have moved of a distance greater than the jump length. $c_0(r,t)$ is therefore close
to the commonly investigated four-point correlation function, at a wavelength related to the inverse jump length.
Then we focus on the spatial correlations between the number of jumps,
which is the spatial correlation of the diffusivity, 
considering the correlation function
\begin{equation}
c_d(r,t) = \frac{\langle n_J(0,t)n_J(r,t) \rangle - \langle n_J(t) \rangle^2}{\langle n_J(t)^2 \rangle- \langle n_J(t) \rangle^2}.
\end{equation}
\begin{figure}[t!]
\begin{center}
\includegraphics*[scale=0.33]{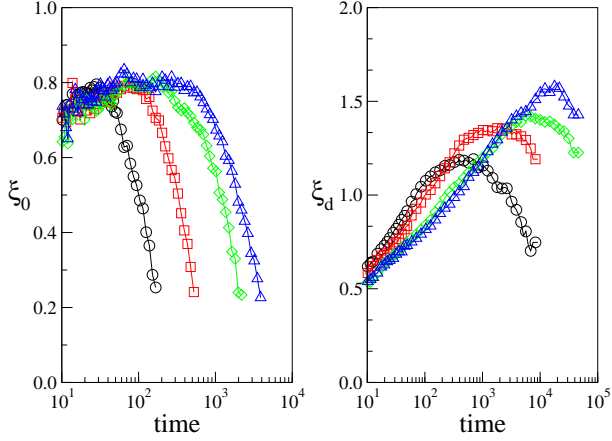}
\end{center}
\caption{\label{fig:spatial}
Spatial correlation lengths. Panel a illustrates the correlation
length of the particles that have performed no jumps, 
panel b the diffusivity correlation length. From left to right:
$T = 0.55, 0.5, 0.46, 0.45$
}
\end{figure}

\begin{figure}[t!]
\begin{center}
\includegraphics*[scale=0.33]{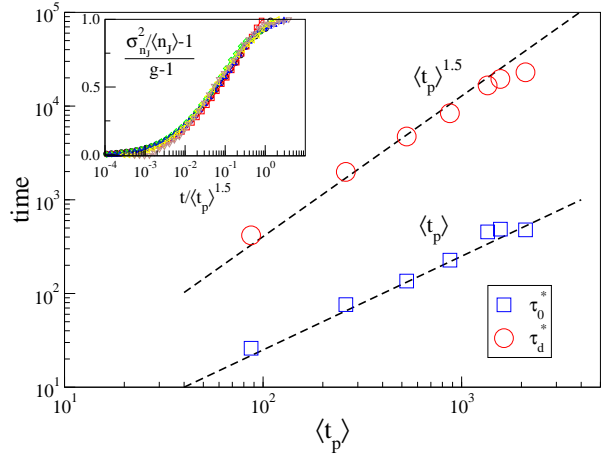}
\end{center}
\caption{\label{fig:temporal}
Temporal correlation lengths. Scaling of the times at which
the correlation lengths $\xi_0$ and $\xi_d$ acquire their
maximum value, with the persistence correlation time, $\<\twf\>$.
The inset shows the rescaling of the data of Fig.~\ref{fig:panelg}a,
for temperatures $T \leq 0.55$.
}
\end{figure}

We find both correlations functions to decay exponentially, with correlation
$\xi_0(t)$ and $\xi_d(t)$, respectively. Their time dependence is illustrated in 
Fig.~\ref{fig:spatial}, for selected temperatures. Both correlation lengths have a maximum as a function of time,
We indicate with $\tau_0^*$ and $\xi^*_0$, and with $\tau_d^*$ and $\xi_d^*$, 
the time of occurrence and the value of the maxima of the two correlation lengths.
As apparent form Fig.~\ref{fig:spatial}, both correlations length are small,
as usual in structural glasses, and increases on cooling, $\xi^*_d$ being much more temperature dependent
than $\xi^*_0$. We characterize the temperature dependence of $\tau_0^*$ and $\tau_d^*$
investigating their scaling with respect to the average persistence time, $\<\twf\>$.
Fig.~\ref{fig:temporal} shows that $\tau_0^* \propto \<\twf\>$, 
in agreement with previous results suggesting that the time of the maximum
of the dynamical heterogeneities scale as the relaxation time.
Conversely, we approximately find $\tau_d^* \propto \<\twf\>^{1.5}$. 
We note that the relation between $\tau_d^*$ and $\<\twf\>$ is model dependent, 
as for instance we observe $\tau_d^* \propto \<\twf\>$ in the Kob--Andersen lattice gas model.
Since $\tau_d^*$ controls the diffusivity correlations,
we expect it to also control the approach of the diffusivity
distribution to its asymptotic Gaussian shape,
and thus to be the time scale at which the 
variance to mean ratio $\sigma^2_{n_J}/\<n_J\>$
reaches its asymptotic value $g$, as in Fig.~\ref{fig:panelg}a.
Indeed, the data of Fig.~\ref{fig:panelg}a are successfully
rescaled when normalized and plotted versus
$t/\tau_d^*$, as in Fig.~\ref{fig:temporal}(inset).

The study of the time evolution of the diffusivity distribution and of the correlation
between the single particle diffusivities allows to identify
a new relaxation timescale, $\tau_d^*$. This grows faster than
the persistence correlation time on cooling.
The emerging physical scenario is as follows:
the relaxation time of the system, as measured
from the decay of scattering correlation functions,
is essentially determined by $\<\twf\>$, as in Eq.\ref{eq:t_l}.
However, on this time scale the diffusivities of the particles
are spatially correlated, and the two dynamical phases 
are still coexisting, as the diffusivity distribution has not acquired
its asymptotic normal shape. It is only on a time of the order of $\tau_d^*$
that all correlations are lost. On this time scale the diffusivity distribution 
has a Gaussian shape, and the diffusivities are spatially uncorrelated.

\section*{Discussion}
Our results show that the dynamics of supercooled liquids 
is characterized by the temporary coexistence of two
phases with different diffusivities, one can reveal
by describing the intermittent particle motion
within the CTRW approach. The presence of
these two phases is related to breakdown of the SE relation,
that also fixes the variance to mean ratio of diffusivity distribution
in the long time limit.
The dynamical phase transition is characterized
by a time scale $\tau_d^*$, 
which is that after which the single particle diffusivities 
are both temporally and spatially uncorrelated. 
The temperature dependence of this time scale
is model dependent, and we have observed it to scale as 
$\<\twf\>^{1.5}$ in the KALJ mixture. 
This result indicates that the 
mean squared displacement grows linearly in time for $t > \<\twf\>$,
while the displacement distribution becomes Gaussian on a larger timescale, for $t > \tau_d^*$.
Accordingly, in between these two time scales the dynamics of the system is
Fickian but not Gaussian~\cite{Granick, Granick14, WeeksScience, Kob07}.

The clear identification of different dynamical phases might also allow
to clarify the debated existence of correlation between the
structural and the dynamical properties of supercooled liquids
\cite{Tanaka,Harrowell2004,Harrowell2006,Hocky2014,Speck2012,Malins}.
Indeed, these correlations have been looked for arbitrarily
dividing the particles in a slow and in a fast phase, introducing
a threshold on the particles' displacements, and then considering
how these phases are related to structural properties, such as V\"oronoi volume, 
local order parameters, local elastic constants, or excess entropy. 
Our results suggest that the slow and the fast phase should
correspond to phases with a high and a small diffusion coefficient, 
we have shown to be unambiguously identified. 

\section*{Methods}
We have performed NVT molecular dynamics 
simulations\cite{LAMMPS} of a $N = 10^3$
standard Kob--Andersen $80$:$20$ ($a$:$b$) binary Lennard--Jones (LJ) mixture\cite{KA1}.
Particles of species $i$ and $j$
interact via a LJ potential with energy scale $\varepsilon_{ij}$
and length scale $\sigma_{ij}$. Values are set as follow:
$\varepsilon_{aa} = 1.0; \sigma_{aa} = 1.0;
\varepsilon_{ab} = 1.5; \sigma_{aa} = 0.8;
\varepsilon_{bb} = 0.5; \sigma_{aa} = 0.88$.
Particles have the same mass $m$.
$\varepsilon_{aa}$, $\sigma_{aa}$ and $m$ are our units of
energy, length and mass.
For each temperature, we have first performed $200$ simulations
to obtain a smooth mean square displacement, from which we have
extracted the Debye--Waller factor $\<u^2\>(T)$ as in Ref.\citenum{Leporini}.
We have then performed other $100$ simulations to investigate the statistical
features of the cage--jump motion as in Ref.\citenum{PastoreSM}:
we associate to each particle, at each time $t$,
the fluctuations $S^2(t)$ of its position computed
over the interval $[t-10t_b:t+10t_b]$, with
$t_b$ ballistic time. The trajectory of each particle
is segmented in cages and jumps, considering a
particle to exit (enter) a cage 
as $S^2(t)$ becomes smaller (larger) than $\<u^2\>$.
This procedure gives access to the duration of each cage, $\tw$, and to duration $\Delta t_j$ and length 
$\Delta r_J$ of each jump. An analogous study has been performed for a $50:50$ two dimensional
mixture of particles interacting via a Harmonic potential\cite{PastoreSM}. 
In the case of the KA lattice kinetically constrained lattice glass model\cite{KAKC},
each particle movement is considered to be a jump. 

\section*{Appendix -- CTRW}
The Continuous Time Random Walk (CTRW) approach describes particle motion in supercooled liquids as a stationary
isotropic walk process\cite{Montroll}.
The temporal features of this process are fixed by the distribution $F(\twf)$ of the persistence time $\twf$,
which is the time particles wait before making their first step as measured from an arbitary $t= 0$ reference time. 
$F(\twf)$ is related to the distribution of the time $\tw$ particles spend in their cages through the Feller relation\cite{Lax, Feller},
$F(\twf) = \<\tw\>^{-1} \left( 1- \int_0^{\twf} P(\tw) d\tw \right)$.
The spatial features are fixed by the distribution of the step size $P(\Delta r)$. 
The walk is assumed to be separable as no correlations
between $\Delta r$ and $\twf$ are considered. The relaxation dynamics
is monitored by the persistence correlation function\cite{PRL2011, berthier_epl2005, Chandler_SE, Hedges2007, Chaudhuri} 
$p(t)=1-\int_{\twf=0}^{t} {F(\twf) d\twf}$, that equals the fraction of particles that has not moved up
to time $t$. Accordingly the relaxation time $\tau$, $p(\tau) = 1/e$, scales as $\<\twf\>$;
conversely, the diffusivity $D$ scales as the number of steps per unit time, $D \propto \<\Delta r^2\>/\<\tw\>$.
While the CTRW approach assumes the waiting times of different particles to be uncorrelated,
this assumption can be easily relaxed to capture the temporal heterogeneities of the dynamics.
Indeed, if temporal correlations involve groups of $M$ particles, then the 
fluctuation of the persistence of a $N$ particle systems scales as $\chi(t) = N\left( \<p(t)^2\>-\<p(t)\>^2 \right) \propto M\left[ \<p(t)\>(1-\<p(t)\>) \right]$,
while its maximum value scales as $\chi^* \propto M$.

\section*{Aknowledgments}
We thank H. Tanaka, A. de Candia, A. Fierro and A. Piscitelli for discussions,
and acknowledge financial support 
from MIUR-FIRB RBFR081IUK, 
from the SPIN SEED 2014 project {\it Charge separation and charge transport in hybrid solar cells},
and from the CNR--NTU joint laboratory {\it Amorphous materials for energy harvesting applications}.

\section*{Author contributions}
R.P., A.C. and M.P.C. conceived the project,
R.P. and M.P.C carried out simulations and analyzed the data,
R.P., A.C. and M.P.C. wrote the paper.

\section*{Additional Information}
The authors declare no competing financial interests.


\begin{thebibliography}{30} 

\bibitem{DHbook}
Berthier, L., Biroli, G., Bouchaud, J-P., Cipeletti, L. \& van Saarloos, W. {eds.} {\it Dynamical Heterogeneities in Glasses, Colloids, and Granular Media}, 
 (Oxford University Press, Oxford, 2011).

\bibitem{Kob97}
Kob, W., Donati, C., Plimpton, S.J., Poole, P.H. \& Glotzer, S.C. Dynamical heterogeneities in a supercooled Lennard-Jones liquid.
 {\it Phys. Rev. Lett.} {\bf 79}, 2827 (1997).

\bibitem{Donati98}
Donati, C. et al. 
Stringlike cooperative motion in a supercooled liquid. {\it Phys. Rev. Lett.} {\bf 80}, 2338 (1998).

\bibitem{Donati99}
Donati, C., Glotzer S.C., \& Poole, P.H..
 Growing spatial correlations of particle displacements in a simulated liquid on cooling toward the glass transition. 
 {\it Phys. Rev. Lett.} {\bf 82}, 5064 (1999).

\bibitem{Kob07} Chaudhuri, P., Berthier, L., \& Kob, W. 
Universal nature of particle displacements close to glass and jamming transitions.  {\it Phys. Rev. Lett.} {\bf 99}, 060604 (2007).


\bibitem{Schmidt1991}
Schmidt--Rohr, K. \& Spiess, H.W.
Nature of nonexponential loss of correlation above the glass transition investigated by multidimensional NMR.
{\it Phys. Rev. Lett.} {\bf 66}, 3020 (1991).

\bibitem{Glotzer}
Glotzer, S.C.
Spatially heterogeneous dynamics in liquids: insights from simulation.
{\it J. Non--Cryst. Solids} {\bf 274}, 342 (2000).

\bibitem{Ediger}
Ediger, M.D.
Spatially heterogeneous dynamics in supercooled liquids.
{\it Annu. Rev. Phys. Chem.} {\bf 51}, 99 (2000).

\bibitem{Chandler_Science}
Hedges, L.O., Jack, R.L., Garrahan J.P. \&  Chandler, D.
Dynamic Order-Disorder in Atomistic Models of Structural Glass Formers.
{\it Science} {\bf 323}, 1309 (2009).

\bibitem{Cammarota2010}
Cammarota, C. et al.
Phase-Separation Perspective on Dynamic Heterogeneities in Glass-Forming Liquids.
{\it Phys. Rev. Lett.} {\bf 105}, 055703 (2010).
\bibitem{Tanaka}
Tanaka, H., Kawasaki, T., Shintani, H., \& Watanabe, K.
Critical-like behaviour of glass-forming liquids. 
{\it Nature Materials} {\bf 9}, 324 (2010).

\bibitem{WeeksScience}
Weeks, E.R., Crocker, J.C., Levitt, A.C., Schofield, A. \& Weitz D.A.
Three-Dimensional Direct Imaging of Structural Relaxation Near the Colloidal Glass Transition.
{ \it Science} {\bf 287}, 627 (2000).



\bibitem{Granick} 
Wang, B., Kuo, J., Bae, S.C., \& Granick, S.
When Brownian diffusion is not Gaussian.
{\it Nature Materials} {\bf 11}, 481 (2012).

\bibitem{Sengupta}
Sengupta, S., \& Karmakar, S.
Distribution of diffusion constants and Stokes-Einstein violation in supercooled liquids.
{\it J. Chem. Phys.} {\bf 140}, 224505 (2014).

\bibitem{KA1}
Kob, W. \& Andersen, H.C.
Scaling Behavior in the $\beta$--Relaxation Regime of a Supercooled Lennard-Jones Mixture.
{\it Phys. Rev. Lett.} {\bf 73}, 1376 (1994).

\bibitem{KA2}
Kob, W. \& Andersen, H.C.
Testing mode--coupling theory for a supercooled binary Lennard-Jones mixture I: The van Hove correlation function.
{\it Phys. Rev. E} {\bf 51}, 4626 (1995).
 
\bibitem{KA3}
Kob, W. \& Andersen, H.C.
Testing mode-coupling theory for a supercooled binary Lennard-Jones mixture. II. Intermediate scattering function and dynamic susceptibility.
{\it Phys. Rev. E} {\bf 52}, 4134 (1995).

\bibitem{PastoreSM}
Pastore, R., Coniglio, A. \& Ciamarra, M.P.
From cage--jump motion to macroscopic diffusion in supercooled liquids.
{\it Soft Matter} {\bf 10}, 5724 (2014).
 
\bibitem{KAKC}
Kob, W. \& Andersen, H.C.,
Kinetic lattice--gas model of cage effects in high--density liquids and a test of mode-coupling theory of the ideal-glass transition.
{\it Phys. Rev. E}  {\bf 48}, 4364 (1993).


\bibitem{PRL2011}
Pastore, R., Ciamarra, M.P., de Candia A. \& Coniglio, A. 
Dynamical Correlation Length and Relaxation Processes in a Glass Former.
{\it Phys. Rev. Lett.} {\bf 107}, 065703 (2011).

\bibitem{Inter1}
Appignanesi, G.A., Fris, J.A.R., Montani, R.A. \& Kob, W.
Democratic Particle Motion for Metabasin Transitions in Simple Glass Formers.
{\it Phys. Rev. Lett.} {\bf 96}, 057801 (2006).

\bibitem{Inter2}
Vogel, M., Doliwa, B., Heuer, A. \&  Glotzer, S.C.
Particle rearrangements during transitions between local minima of the potential energy landscape of a binary Lennard-Jones liquid.
{\it J. Chem. Phys.} {\bf 120}, 4404 (2004).

\bibitem{Inter3}
Vallee, R.A.L., van der Auweraer, M., Paul, W. \& Binder, K. 
Fluorescence Lifetime of a Single Molecule as an Observable of Meta--Basin Dynamics in Fluids Near the Glass Transition.
{\it Phys. Rev. Lett.} {\bf 97}, 217801 (2006).

\bibitem{Inter4}
Fris, J.A.R., Appignanesi, G.A. \& Weeks, E.R.
Experimental Verification of Rapid, Sporadic Particle Motions by Direct Imaging of Glassy Colloidal Systems.
{\it Phys. Rev. Lett.} {\bf 10}, 065704 (2011).

\bibitem{Inter5}
Vollmayr-Lee, K. 
Single particle jumps in a binary Lennard-Jones system below the glass transition.
{\it J. Chem. Phys.} {\bf 121}, 4781 (2004).

\bibitem{Makse2009}
Carmi, S., Havlin, S., Song, C., Wang K. \& Makse, H.A.
An energy landscape network approach to the glass transition.
{\it J. Phys. A: Math. Theor.} {\bf 42}, 105101 (2009).

\bibitem{SM15}
Pastore, R., Pica Ciamarra, M., Pesce, G. \& Sasso, A.
Connecting short and long time dynamics in hard--sphere--like colloidal glasses.
{\it Soft Matter} {\bf 11}, 622 (2015).

\bibitem{berthier_epl2005}
Berthier, L., Chandler, D. \& Garrahan, J.P.
Length scale for the onset of Fickian diffusion in supercooled liquids.
{\it Europhys. Lett.} {\bf 69} 320 (2005).

\bibitem{Chandler_SE}
Jung, Y., Garrahan, J.P. \& Chandler D.,
Excitation lines and the breakdown of Stokes--Einstein relations in supercooled liquids.
{\it Phys. Rev. E} 69, 061205 {\bf 2004}.

\bibitem{Debenedetti}
Debenedetti, P.G. \& Stillinger, F.H.
Supercooled liquids and the glass transition.
{\it Nature} {\bf 410}, 259 (2001).


\bibitem{Hedges2007}
Hedges, L.O., Maibaum, L., Chandler, D. \&  Garrahan, J.P.
Decoupling of exchange and persistence times in atomistic models of glass formers.
{\it J. Chem. Phys.} {\bf 127}, 211101 (2007).

\bibitem{Feller49}
Feller, W., 
Fluctuation theory of recurrent events. 
{\it Trans. Amer. Math. Soc.} {\bf 67}, 98 (1949).

\bibitem{FellerBook}
Feller, W.
{\it An Introduction to Probability Theory and its Applications} (Wiley, New York, 1971), Vol. 2, 2$^{\rm nd}$ ed.

\bibitem{Lax}
Lax, M. \& Scher, H.
Renewal Theory and ac Conductivity in Random Structures.
{\it Phys. Rev. Lett.} {\bf 39}, 781 (1977).



\bibitem{Granick14}
Guan, J., Wang, B. \& Granick, S.
Even Hard--Sphere Colloidal Suspensions Display Fickian Yet Non-Gaussian Diffusion.
{\it ACS nano} {\bf 8}, 331 (2014).

\bibitem{Harrowell2004}
Widmer-Cooper, A., Harrowell, P. and Fynewever, H.,
How Reproducible Are Dynamic Heterogeneities in a Supercooled Liquid?,
{\it Phys. Rev. Lett.} {\bf 93}, 135701 (2004).

\bibitem{Harrowell2006}
Widmer--Cooper A. \& Harrowell, P.
Free volume cannot explain the spatial heterogeneity of Debye--Waller factors in a glass-forming binary alloy.
{\it J. Non-Cryst. Solids} {\bf 352}, 5098 (2006).

\bibitem{Hocky2014}
Hocky, G.M., Coslovich, D., Ikeda, A., \& Reichman, D.R.
Correlation of Local Order with Particle Mobility in Supercooled Liquids Is Highly System Dependent
{\it Phys. Rev. Lett.} {\bf 113}, 157801 (2014).

\bibitem{Speck2012}
Speck, T., Malins, A., \& Royall, C.P.
First--Order Phase Transition in a Model Glass Former: Coupling of Local Structure and Dynamics.
{\it Phys. Rev. Lett.} {\bf 109}, 195703 (2012).



\bibitem{Malins}
Malins, A. Williams, S.R., Eggers, J., \& Royall, C.P.
Identification of structure in condensed matter with the topological cluster classification, 
{\it J. Chem. Phys.} {\bf 139}, 234506 (2013).

\bibitem{LAMMPS}
Plimpton, S.
Fast Parallel Algorithms for Short-Range Molecular Dynamics.
{\it J. Comp. Phys.} {\bf 117}, 1 (1995).


\bibitem{Leporini}
Larini, L., Ottochian, A., De Michele C. \& Leporini, D.
Universal scaling between structural relaxation and vibrational dynamics in glass-forming liquids and polymers.
{\it Nature Physics}  {\bf 4}, 42 (2007). 

\bibitem{Montroll}
Montroll, E.W. \& Weoss, G.H. 
{\it J. Math. Phys.} {\bf 6}, 167 (1965).

\bibitem{Feller}
Tunaley, J.K.E.
Theory of ac Conductivity Based on Random Walks.
{\it Phys. Rev. Lett.} {\bf 33}, 1037 (1974).

\bibitem{Chaudhuri}
Chaudhuri, P., Sastry,  S. \& Kob, W.
Tracking Heterogeneous Dynamics During the $\alpha$ Relaxation of a Simple Glass Former.
{\it Phys. Rev. Lett.} {\bf 101}, 190601 (2008).



\end{thebibliography}
\end{document}